\documentclass[11pt]{article}
\usepackage{cospar}
\usepackage{url}
\usepackage{t1enc}
\usepackage[latin1]{inputenc}
\usepackage[english]{babel}

\newcommand{\gr}{$^{\circ}$}

\newcommand{\refbib}{\hangindent\parindent\noindent}

\title{MIRAX: A BRAZILIAN X-RAY ASTRONOMY SATELLITE MISSION}

\author{J. Braga\address{Instituto Nacional de Pesquisas Espaciais
(INPE), C.P. 515, S.J. Campos, SP 12201-970, Brazil},
R. Rothschild\address{University of California San Diego (UCSD), 9500
Gilman Dr., La Jolla, CA 92093-0424, USA}, J. Heise\address{Space
Research Organization Netherlands (SRON), Sorbonnelaan 2 3485 CA
UTRECHT, The Netherlands}, R. Staubert\address{Institut f\"ur
Astronomie und Astrophysik (IAAT), Sand 1, T\"ubingen 72076, Germany},
R. Remillard\address{Massachusetts Institute of Technology (MIT), 70
Vassar St., Cambridge, MA 02139, USA}, F. D'Amico$^1$, F.
Jablonski$^1$, W. Heindl$^2$,
J. Matteson$^2$, E. Kuulkers\address{European Space Research \&
Technology Center (ESTEC/ESA), Keplerlaan 1, Postbus 299, 2200 AG
Noordwijk, The Netherlands}, J. Wilms$^4$, E. Kendziorra$^4$}

\begin{document}

\maketitle

\begin{abstract}
We describe the ``Monitor e Imageador de Raios-X'' (MIRAX), an X-ray
astronomy satellite mission proposed by the high energy astrophysics
group at the National Institute for Space Research (INPE) in Brazil to
the Brazilian Space Agency. MIRAX is an international collaboration
that includes, besides INPE, the University of California San Diego,
the University of Tübingen in Germany, the Massachusetts Institute of
Technology and the Space Research Organization Netherlands. The
payload of MIRAX will consist of two identical hard X-ray cameras (10
-200 keV) and one soft X-ray camera (2-28 keV), both with angular
resolution of $\sim$5-7 arcmin. The basic objective of MIRAX is to carry
out continuous broadband imaging spectroscopy observations of a large
source sample ($\sim$9 months/yr) in the central Galactic plane
region. This will allow the detection, localization, possible
identification, and spectral/temporal study of the entire history of
transient phenomena to be carried out in one single mission. MIRAX
will have sensitivities of $\sim$ 5 mCrab/day in the 2-10 keV band ($\sim$ 2
times better than the All Sky Monitor on Rossi X-ray Timing Explorer)
and 2.6 mCrab/day in the 10-100 keV band ($\sim$ 40 times better than the
Earth Occultation technique of the Burst and Transient Source
Experiment on the Compton Gamma-Ray Observatory). The MIRAX spacecraft
will weigh about 200 kg and is expected to be launched in a
low-altitude ($\sim$600 km) circular equatorial orbit around 2007/2008.
\end{abstract}

\section*{INTRODUCTION}

The ``Monitor e Imageador de Raios-X'' (MIRAX) is a high-energy
astrophysics satellite mission which is part of the space science
microsatellite program at the National Institute for Space Research
(INPE) in Brazil. MIRAX has been selected to be the astrophysics
mission within this program and has been proposed to the Brazilian
Space Agency (AEB). Since the Brazilian astronomical community is
mostly devoted to the fields of optical and radio astronomy, the
development and operation of MIRAX is expected to have a major impact
on Brazilian science through the opening of a new observation window
for astrophysical research.  The MIRAX project has strong
international partnership. The University of California in San Diego
(UCSD) will provide the hard X-ray detectors and participate in the
design of the hard X-ray cameras; the Space Research Organization
Netherlands (SRON) is expected to provide the soft X-ray imager; the
Institut für Astronomie und Astrophysik of the University of Tübingen
(IAAT) will provide the on-board computer and participate in software
development; and the Massachusetts Institute of Technology (MIT) will
participate in software development for data acquisition, analysis,
storage and distribution.  The main scientific goal of MIRAX is the
nearly continuous (9 months per year) , broad-band (2 to 200 keV),
high-resolution ($\sim$5-7 arcminutes) monitoring of a specific large
region of the sky that is particularly rich of X-ray sources (a 76\gr\
x 44\gr\ field centered on the Galactic center and oriented along the
Galactic plane). This will not only provide an unprecedented
monitoring of the X-ray sky through simultaneous spectral observations
of a large number of sources, but will also allow the detection,
localization, possible identification, and spectral/temporal study of
the entire history of transient phenomena to be carried out in one
single mission. During the $\sim$3 months/year when the Sun will be
crossing the central Galactic plane, MIRAX will be pointed to other
rich fields such as the Magellanic Clouds and the Cygnus and
Vela/Centaurus regions. MIRAX will be able to contribute to the study
of a variety of phenomena and objects in high energy astrophysics,
especially in the so far poorly explored non-thermal domain of hard
X-ray observations. With the planned continuous monitoring approach,
MIRAX will address key issues in the field of X-ray variability such
as black hole state transitions and early evolution, accretion torques
on neutron stars (especially through monitoring of X-ray pulsars),
relativistic ejections on microquasars and fast X-ray novae. MIRAX
will also be able to contribute to Gamma-Ray Burst (GRB) astronomy,
since it is expected that $\sim$1 GRB will be detected per month in
MIRAX's field-of-view (FOV). MIRAX will not only provide positions of
GRBs with an accuracy a few arcminutes but will also obtain broadband
X-ray spectra of the bursts and possibly their X-ray afterglows. MIRAX
instruments are expected to be assembled in a dedicated small
($\sim$200 kg) satellite to be launched in a low altitude, equatorial
circular orbit around 2007/2008. Table 1 shows the baseline parameters
of MIRAX. In comparison with the Burst Alert Telescope (BAT) on the
Swift mission, expected to be launched in 2003, MIRAX has a smaller
detector area in the hard X-ray range (factor of $\sim$7) but a higher
angular resolution (factor of 2.3). The main advantage of MIRAX over
BAT is the continuous viewing approach for the study of transient
phenomena and variability.

\bigskip
\noindent Table 1. MIRAX baseline parameters

\begin{tabular}{|l|l|l|}
\hline
\multicolumn{3}{|l|}{\bf Mission and spacecraft parameters} \\
\hline  
Mass & \multicolumn{2}{|l|}{$\sim$200 kg (total), $\sim$100 kg
(payload)} \\
Power & \multicolumn{2}{|l|}{$\sim$240 W (total), $\sim$90 W
(payload)} \\
Orbit & \multicolumn{2}{|l|}{equatorial, circular, $\sim$600 km} \\
Telemetry & \multicolumn{2}{|l|}{S-band (2200-2290 MHz), $\sim$1.5
Mbps downlink} \\ 
Launch & \multicolumn{2}{|l|}{2007/2008 by Brazilian VLS (``Veículo
Lançador de Satélites'')} \\
\hline 
\bf Instrument parameters & \bf Hard X-ray Imager (CXD) & \bf Soft X-ray
Imager (CXM) \\
\hline
Energy range & 10-200 keV & 2-28 keV \\
Angular resolution & 7.5 arcmin & 5 arcmin \\
Localization & < $1$ arcmin (10$\sigma$ source) & $<$ 1 arcmin
(10$\sigma$ source) \\
Field-of-view & 58\gr\ x 26\gr\ FWHM along the GP & 20\gr\ x 20\gr\
FWHM \\
Spectral resolution & $<$ 5 keV @ 60 keV & 1.2 keV @ 6 keV \\
Time resolution & $<$ 10 $\mu$s & 122 $\mu$s \\
Sensitivity & $<$ 2.6 mCrab (1 day, 5$\sigma$) & $<$ 5 mCrab (1 day,
5$\sigma$) \\
Detector area & 2 x 360 cm$^2$ & 650 cm$^2$ \\
\hline
\end{tabular}

\section*{MIRAX INSTRUMENTS}

In the current planned configuration, the payload will consist of a
set of two hard X-ray cameras (CXD - ``Câmera de Raios-X Duros'') and
one soft X-ray camera (CXM - ``Câmera de Raios-X Moles'').  Both imagers
will employ the technique of coded-aperture imaging (Dicke, 1968;
Skinner, 1984; Caroli et al., 1987; Braga, 1990), which has been
highly successful on X-ray satellite instruments such as Spacelab
2/XRT (Willmore et al., 1984), GRANAT/ART-P (Sunyaev et al., 1990),
RXTE/ASM (Levine et al., 1996), BeppoSAX/WFC (Jager et al., 1997),
Kvant/COMIS-TTM (In 't Zand, 1992), and especially GRANAT/SIGMA
(Roques et al., 1990; Paul et al., 1991; Bouchet et al., 2001), as
well as on balloon experiments such as GRIP-2 (Schindler et al., 1997)
and EXITE (Garcia et al., 1986; Braga, Covault and Grindlay, 1989).

\subsection*{The Hard X-Ray Cameras}

The CXDs will be built in collaboration with the Center for
Astrophysics and Space Science (CASS) of UCSD and will operate from 10
to 200 keV. The detector plane will be a 3 x 3 array of
state-of-the-art CdZnTe crossed-strip detector modules with 0.5 mm
spatial resolution developed at CASS, with a total area of 360
cm$^2$. Each detector module is a 2 x 2 array of 32 mm x 32 mm x 2mm
thick CZT detectors. The detectors will be surrounded by an active
plastic scintillator shield and by a passive Pb-Sn-Cu graded shield. A
315 mm x 275 mm Tungsten coded-mask with 1.3 mm-side square cells (0.5
mm-thick) will be placed 600 mm away from the detector to provide
images with 7'30" angular resolution. The basic pattern of the mask
will be a 139 x 139 Modified Uniformly Redundant Array (MURA -
Gottesman and Fenimore, 1989; Braga et al., 2002), which will allow
for full shadowgrams to be cast on the position-sensitive detector
area and will provide no source ambiguity problems in the fully-coded
field-of-view (FCFOV). A sketch of the CXD is shown in Figure 1.  The
pointing axes of the two CXDs will be offset by an angle of 29\gr\ in
order to provide a uniform sensitivity over a 39\gr\ FCFOV in one
direction; the perpendicular direction will have a 6\gr 12' FCFOV. In
such a configuration the FWHM FOV is 58\gr\ x 26\gr. During the
observations of central Galactic Plane, the wider direction of the FOV
will be aligned with the GP. Figure 2 shows the fractional coded-area
(considering the two cameras) as a function of angle, for the
direction aligned with the GP.

\subsection*{The Soft X-Ray Camera}

The CXM, expected to be provided by SRON, is the spare flight unit of
the Wide Field Cameras (WFCs - Jager et al. 1997) of the
recently-terminated BeppoSAX mission (Boella et al. 1997), and will
operate from 1.8 to 28 keV. The CXM will have a 5' angular resolution
in a 20\gr\ x 20\gr\ FWHM FOV. The addition of the WFC to the MIRAX
payload will provide soft X-ray spectral coverage which will be
extremely important for the study of the several classes of sources in
the MIRAX FOV. Furthermore, the excellent perfomance of the WFCs on
BeppoSAX brings to MIRAX an instrument that has already been tested
and used successfully in orbit with very little degradation on a time
scale of several years.  A preliminary design of the MIRAX spacecraft
is show on Figure 3. The CXM is mounted on top of the two CXDs, which
are offset by 29\gr. A star camera with an Active Pixel Sensor (APS),
currently being developed at INPE, is placed in the space between the
two CXDs.

\subsection*{The Flight Computer}

The instruments on the payload of MIRAX will send data to a Central
Electronis Unit (CEU), which will be the data and command interface
between the imagers and the spacecraft.  The CEU will receive and
process data from the three cameras, select the "good" events
according to a variety of criteria (thresholds, shield vetoes,
calibration source events, etc.) and build the telemetry packets. The
processing at the CEU will include determination of position, energy
and depth of the X-ray interactions in the CZT detectors, as well as
time tagging. The data packets will then be sent to the MIRAX
spacecraft computer for transmission to the ground. The CEU will be
developed by the IAAT with collaboration from INPE. The IATT has
extensive experience on space missions and a strong heritage in flight
computers.  

\subsection*{Sensitivity} 

The MIRAX hard X-ray sensitivity can be estimated based on the
expected background level in the low-orbit environment, which is about
200 cts/s for a single imager. The internal component is calculated
from balloon flights of prototype CZT detectors launched from Fort
Sumner, NM. The Crab nebula plus pulsar photon count rate will be
$\sim$120 cts/s. Taken the approximate total contribution of sources
in the primary MIRAX FOV (central GP) to be about 1 Crab, the MIRAX
sensitivity is expected to better than 2 x 10$^{-5}$ photons cm$^{-2}$
s$^{-1}$ keV$^{-1}$ at 100 keV (5 $\sigma$) , or $\sim$2.6 mCrab/day
in the 10-100 keV range (approximately 40 times better than the Earth
Occultation technique of the Burst and Transient Source Experiment on
CGRO). Techniques for background reduction for CZT strip detectors are
being developed, especially involving vetoing of multiple,
non-contiguous events (expected to come from particle-induced showers
within the surrounding material) and low-energy interactions deep in
the detector, which are produced by photons incident from the bottom
of the detector plane. The CXDs will have a one-year ``survey''
sensitivity, considering a conservative systematics limit of 0.1\% of
background, of about 10-11 erg cm$^{-2}$ s$^{-1}$ in the 10-50 keV
band. This is $\sim$20 times better than what was achieved by the HEAO
1 A-4 instrument, which carried out the only hard X-ray survey to date
(Levine et al., 1984).  For the low energy range, the soft X-ray
imager will have approximately the sensitivity of the WFCs on
BeppoSAX, which is better than 5 mCrab/day in the 2-10 keV band
(approximately 2 times better than the All Sky Monitor on RXTE).

\section*{SPACECRAFT AND MISSION OPERATIONS}

The MIRAX spacecraft will be based on the satellite bus already
developed at INPE for the French-Brazilian microsatellite (FBM)
mission, expected to be launched in 2005. The platform employs a
3-axis attitude stabilization system with 2 star trackers, a sun
sensor and a magnetometer. Torque rods and reaction wheels will be
used as attitude controlers. The MIRAX payload will have no moving
parts and a mass of $\sim$100 kg, while the total spacecraft mass is
expected to be under 200 kg. There will be no propulsion and the
pointing will be inertial. The pointing precision will be 0.5\gr, with
36"/hr stability (1/10 of the image pixel) and 20" attitude
knowledge. The power consumption of the payload will be between 88 and
96 W, depending on the final configuration and the CEU requirements,
and the total power of the satellite will be around 240~W.  The MIRAX
mission duration is required to be 2 years, with a possible extension
to 5 years. A ground station at Natal, Brazil, operated by INPE, will
be assembled. Possibly, a second ground station in Kenya will be
available. The space operation S-band (2200 - 2290 MHz) will be used
for downlink and command uplink. It is expected that downlink data
rates up to $\sim$2 Mbits/s will be possible to reach, depending on
the modulation and on coordination with other satellites. Our current
estimates indicate that a rate of 1.5 Mbits/s will be enough to dump
all the data with no compression if we use one station.  MIRAX is
expected to be launched by 2007/2008 by the Brazilian satellite
launcher VLS, in case it is tested successfully and is officially
considered a reliable launcher by the Brazilian Space Agency. In case
a VLS is not available, other possibilities will be considered, such
as a Pegasus launch or as a piggy-back payload on larger launchers.
MIRAX data will be 100\% available to the community
immediately. Databases will be setup at the missions centers in Brazil
(INPE) and at UCSD. The database will also be available at HEASARC
(Goddard Space Flight Center).  Specific webpages with several data
products will be available.

\section*{REFERENCES}

\refbib Boella, G., R. C. Butler, G. C. Perola, et al., BeppoSAX, the wide
band mission for X-ray astronomy, {\it
Astr. Astrophys. Suppl. Series}, {\bf 122}, 299-307, 1997. 

\refbib Bouchet, L., J. P. Roques, J. Ballet, A. Goldwurm, and J. Paul, The
SIGMA/GRANAT Telescope: Calibration and data reduction, {\it
Astrophys. J.}, {\bf 548}, 990-1009, 2001.

\refbib Braga, J., Técnicas de Imageamento em Raios-X Duros com Máscaras
Codificadas, {\it PhD Thesis}, University of São Paulo, Brazil, 1990.

\refbib Braga, J., C. Covault, and J. Grindlay, Calibration and performance of
the Energetic X-Ray Imaging Telescope Experiment, {\it IEEE
Trans. Nucl. Sci.}, {\bf 36}, 871-875, 1989.

\refbib Braga, J., F. D'Amico, T. Villela, J. Mejía, R. A. Fonseca, and
E. Rinke, Development of the imaging system of the balloon-borne
gamma-ray telescope Máscara Codificada (MASCO), {\it
Rev. Sci. Instr.}, {\bf 73}(10), 3619-3628, 2002.

\refbib Caroli, E., J. B. Stephen, G. Di Cocco, L. Natalucci, and
A. Spizzichino, Coded aperture imaging in X- and gamma ray astronomy,
{\it Space Sci. Rev.}, {\bf 45}, 349-403, 1987.

\refbib Dicke, H. R., Scatter-hole cameras for X-rays and Gamma rays,
{\it Astrophys. J.}, {\bf 153}, L101-L106, 1968.

\refbib Garcia, M., J. Grindlay, R. Burg, S. Murray, and J. Flanagan,
Development of the EXITE detector: a new imaging detector for 20-300
keV astronomy, {\it IEEE Trans. Nucl. Sci.}, {\bf 33}, 735-740, 1986.

\refbib Gottesman, S.R., and E. E. Fenimore, New family of binary arrays for
coded aperture imaging, {\it Applied Optics}, {\bf 28}(20), 4344-4352,
1989.

\refbib In 't Zand, J., A Coded-Mask Imager as Monitor of Galactic Center
Sources, {\it PhD Thesis}, University of Utrecht, The Netherlands, 1992.

\refbib Jager, R., W. A. Mels, A. C. Brinkman, M. Y. Galama, H. Goulooze,
J. Heise, P. Lowes, J. M. Muller, A. Naber, A. Rook, R. Schuurhof,
J. J. Schuurmans, and G. Wiersma, The Wide Field Cameras onboard the
BeppoSAX X-ray Astronomy Satellite, {\it Astr. Astrophys. Suppl. Series},
{\bf 125}, 557-572, 1997.

\refbib Levine, A. M., H. Bradt, W. Cui, J. G. Jernigan, E. H. Morgan,
H. Edward, R. Remillard, R. E. Shirey, and D. A. Smith, First Results
from the All-Sky Monitor on the Rossi X-Ray Timing Explorer,
{\it Astrophys. J. Lett.}, {\bf 469}, L33-L36, 1996.

\refbib Levine, A., F. L. Lang, W. H. G. Lewin, et al., The HEAO 1 A-4
Catalog of High-Enery X-Ray Sources, {\it
Astrophys. J. Suppl. Series}, {\bf 54}, 581-617, 1984.

\refbib Paul, J., J. Ballet, M. Cantin, B. Cordier, A. Goldwurm, A. Lambert,
P. Mandrou, J. P. Chabaud, M. Ehanno, and J. Lande, Sigma - The hard
X-ray and soft gamma-ray telescope on board the GRANAT space
observatory, {\it Adv. Space Res.}, {\bf 11}, 289-302, 1991.

\refbib Roques, J., J. Paul, P. Mandrou, and F. Lebrun, The Sigma
mission on the GRANAT satellite, {\it Adv. Space Res.}, {\bf 10},
223-225, 1990.

\refbib Shindler, S. M., W. R. Cook, J. Hammond, F. A. Harrison, T. A. Prince,
S. Wang, S. Corbel, and W. A. Heindl, GRIP-2: A sensitive
balloon-borne imaging gamma-ray telescope, {\it Nucl. Instr. Meth.},
{\bf 384}, 425-434, 1997.

\refbib Skinner, G. K., Imaging with coded-aperture masks,
{\it Nucl. Instr. Meth. Phys. Res.}, {\bf 221}, 33-40, 1984.

\refbib Sunyaev, R. A., S. I. Babichencko, D. A. Goganov,
S. R. Tabaldyev, and N. S. Jambourenko, X-ray telescopes ART-P and
ART-S for the GRANAT project, {\it Adv. Space Res.}, {\bf 10}(2),
233-237, 1990.

\refbib Willmore, A., G. Skinner, C. Eyles, and B. Ramsey, A coded
mask telescope for the Spacelab 2 mission, {\it
Nucl. Instr. Meth. Phys. Res.}, {\bf 221}, 284-287, 1984.

\vspace{1cm}

\noindent{\bf Figure captions}

\vspace{0.5cm}
\refbib{\bf Fig. 1.} Exploded diagram of the MIRAX hard X-ray
camera. From left to right, the elements are: coded-mask, coded-mask
support structure, Pb-Sn-Cu passive-shield walls, two structural
flanges, detector modules, plastic scintillator and Pb-Sn-Cu passive
shield.

\vspace{0.5cm}
\refbib{\bf Fig. 2.} The fractional coded area of the two hard X-ray
cameras of MIRAX with the main axes offset by 29\gr. This angle provides
a nearly uniform FCFOV of 39\gr\ along the GP.

\vspace{0.5cm}
\refbib{\bf Fig. 3.} A preliminary view of the MIRAX spacecraft. The
two cameras mounted over the satellite bus are the hard X-ray cameras
(CXDs), whereas the soft X-ray camera is on top of the CXDs. The APS
optical star camera is placed in between the CXDs. The external
dimensions are approximately 1.5 m x 0.7 m x 0.7 m (with the solar
panels folded over the spacecraft).

\vspace{0.5cm}

\end{document}